\documentclass[trackchanges,twocolumn]{aastex701}

\usepackage[utf8]{inputenc}
\usepackage{booktabs}   
\usepackage{tabularx}   
\usepackage{enumitem}   
\usepackage{xcolor}     
\usepackage{array}

\usepackage{siunitx}

\usepackage{natbib}
\usepackage{amsmath}
\usepackage{subfigure}
\usepackage{float,color}
\usepackage{gensymb}
\usepackage{placeins}
\usepackage{multirow}
\usepackage{appendix}

\begin{document}

\title{Leveraging Lensing-driven Early Warnings: Joint Detectability of lensed Gravitational Waves and their LSST Optical Counterparts}

\author[0000-0002-0870-2993]{Sourabh Magare}
\affiliation{Inter-University Centre for Astronomy and Astrophysics, Post Bag 4, Ganeshkhind, Pune - 411007, India}
\email{sourabh.magare@gmail.com}

\author[0000-0002-0991-8438]{Anindya }
\affiliation{Inter-University Centre for Astronomy and Astrophysics, Post Bag 4, Ganeshkhind, Pune - 411007, India}
\email{anindya.ganguly@iucaa.in}

\author[0000-0001-7714-7076]{Anupreeta More}
\affiliation{Inter-University Centre for Astronomy and Astrophysics, Post Bag 4, Ganeshkhind, Pune - 411007, India}
\affiliation{Kavli IPMU (WPI), UTIAS, The University of Tokyo, Kashiwa, Chiba 277-8583, Japan}
\email{anupreeta@iucaa.in}

\author[0000-0001-5318-1253]{Shasvath J. Kapadia}
\affiliation{Inter-University Centre for Astronomy and Astrophysics, Post Bag 4, Ganeshkhind, Pune - 411007, India}
\email{shasvath.kapadia@iucaa.in}

\begin{abstract}
Gravitational lensing of gravitational waves (GWs) can be leveraged to provide early-warning times of $\mathcal{O}(\mathrm{hours})$ to $\mathcal{O}(\mathrm{days})$ prior to the mergers of binary neutron stars (BNSs) and neutron star-black holes (NSBHs). This capability could, in turn, enable electromagnetic (EM) observatories to capture transient emissions surrounding the time of the merger. In this work, we demonstrate the viability of jointly detecting gravitationally lensed BNS and NSBH mergers via GWs during the O5 observing run, alongside their associated kilonova counterparts using the Legacy Survey of Space and Time (LSST). We simulate a population of lensed BNS and NSBH mergers detectable during O5 and utilize current benchmark forward models to generate the corresponding lensed kilonova light curves. We then assess the detectability of these light curves across LSST bands under various exposure limits. We find that among the detectable lensed mergers in O5, approximately 60\% of the BNS kilonova and 10\% of the NSBH kilonova will be observable within a single LSST exposure, with mostly quads dominating the detectable population. Furthermore, we define a detectability index ($\rm DI$), which determines the EM-detectability in optical of lensed GW-detectable BNS and NSBH mergers. We demonstrate the robustness and generalizability of the $\rm DI$ on our lensed BNS and NSBH population. Our work demonstrates that lensing-driven early warnings will be highly applicable in the near future, enabling the LSST to effectively follow up on BNS and NSBH triggers within GW localization areas during the O5 observing run.

\end{abstract}


\section{Introduction}

The joint detection of gravitational waves (GW) from GW170817 by the LIGO-Virgo Collaboration \citep{LIGOScientific:2014pky, 2015CQGra..32b4001A, 2017PhRvL.119p1101A} and the kilonova AT2017gfo by electromagnetic (EM) telescopes \citep{2017Sci...358.1556C, 2017Natur.551...75S} marked the advent of multi-messenger astronomy \citep{2017ApJ...848L..12A, 2017ApJ...848L..13A}.   
This was the first detection of a binary neutron star (BNS) merger and was followed by telescopes that discovered the first-ever confirmed kilonova, AT2017gfo, in NGC 4993.  
This coincidental observation confirmed that the kilonova is a result of BNS mergers, and that neutron star (NS) mergers are the primary sites for heavy-element nucleosynthesis via the r-process \citep{2017Natur.551...67P}. This joint detection imposed vital constraints on several physical theories. It provided upper bounds for graviton mass \citep{PhysRevD.100.104036}, refined bounds on the speed of GWs \citep{2017ApJ...848L..12A}, and placed constraints on Lorentz invariance and the equivalence principle \citep{PhysRevD.97.124023, PhysRevD.97.084040}. Additionally, it also clarified the nature of short GRBs and kilonova, identifying the sources behind these high-energy bursts \citep{2017ApJ...848L..12A}. 

The thermal transient from the BNS merger had long been predicted in the form of a kilonova \citep{1998ApJ...507L..59L, 2010MNRAS.406.2650M}. A kilonova is a non-relativistic outflow heated by the decays of heavy nuclei formed through rapid neutron captures \citep{Freiburghaus_1999}. The total ejected mass from the GW170817 kilonova was inferred to be $\sim 0.05 M_{\odot}$, with at least two components of different opacities \citep{2017ApJ...851L..21V, 2021ARA&A..59..155M}. The significant ejecta mass and broad isotopic distribution of atomic masses, spanning all three r-process abundance peaks \citep{RevModPhys.29.547}, strongly suggest that neutron star mergers are the dominant astrophysical sites for heavy r-process nucleosynthesis. 
Thus, studying kilonova is crucial for understanding cosmic chemical evolution. 
la sources
Further, BNS mergers provide an excellent laboratory for determining the neutron star equation of state (EoS) \citep{2016PhR...621..127L}. The EoS determines the neutron star radius, the maximum supportable mass and tidal deformability, and direct measurements of these quantities 
from astrophysical observations constrain the EoS. For GW170817, \citep{2017ApJ...850L..19M} used the observed EM data of GW170817 to infer that its remnant must have collapsed to a black hole within a timescale of ($\sim 10$ ms to $\sim 1$s, pointing to $M_{\rm TOV}$ of $2.17 M_{\odot}$. \citep{2017ApJ...850L..34B} related this lifetime to the radius, finding $R_{1.4} > 10.68$km. \citep{PhysRevLett.120.261103} constrains the radius of neutron star to $12 <R_{1.4}/km < 13.45$ at $2\sigma$ confidence level. These measurements were further refined by \citep{Nicholl:2021rcr}, who modeled the kilonova light curves using an analytic forward model parameterized by binary properties and the equation of state. 
Furthermore, the kilonova peak brightness can be used to investigate the compactness of the neutron star, estimate the ejected mass and tidal deformability, and identify whether it's a BNS or NSBH \citep{2019EPJA...55...50R, 2020ApJ...889..171K, PhysRevD.111.064022}. 


Despite the groundbreaking nature of GW170817, a notable latency persisted between the initial GW detection and EM follow-up \citep{2017ApJ...848L..12A}. This delay underscores the critical requirement for early-warning systems capable of alerting telescopes before the actual merger, thereby enabling the capture of pre-merger \citep[e.g.,][]{PhysRevLett.108.011102, Most_2020}, merger-phase, and post-merger EM signals \citep{Siegel_2016}. 

Rapid acquisition of optical and ultraviolet spectra would permit the mapping of ejecta mass and velocity distributions, ultimately refining current models of r-process nucleosynthesis \citep{Nicholl_2017}. Similarly, prompt X-ray observations are essential for characterizing the physical state of the post-merger remnant \citep{Siegel_2016}. Furthermore, early-time radio spectroscopy could provide a vital test of theories suggesting that magnetic interactions during BNS coalescences power fast radio bursts \citep{Most2020ElectromagneticPT, Wang_2016}. Ultimately, observing EM emissions in the pre-merger window holds the potential to resolve fundamental, long-standing questions in neutron star astrophysics.

In our earlier work \citep{Magare_2023}, we introduced a novel astrophysical framework capable of providing an early warning of $\mathcal{O}(\mathrm{hours})$ to $\mathcal{O}(\mathrm{days})$ prior to a merger, achieving sub-arcsecond sky localization, provided the BNS or neutron star-black hole (NSBH) system is strongly lensed. This scenario dictates that if we detect gravitational waves (GWs) from a BNS or NSBH event with an EM counterpart, subsequent EM follow-up observations can localize the merger to its host galaxy, as with GW170817. If this host galaxy is found to be strongly lensed, the positions of the galactic images, combined with the redshifts of both the host and the lens galaxy, can be used to predict the arrival times of subsequent lensed images. The anticipated arrival of these future images effectively serves as an early warning for the BNS or NSBH merger, thereby drastically extending the available pre-merger alert window. In our follow-up work \citep{Magare_2026}, we use simulated galaxy-scale lens systems to evaluate the feasibility of this method for early warning predictions of subsequent EM (and GW) counterparts.   


An important metric for assessing the feasibility of lensing-driven early warnings is whether lensed kilonova from BNS or NSBH mergers are detectable with ground-based survey telescopes such as LSST. Most previous studies have focused on estimating the detection rates of lensed GW events, rarely considering the detectability of their EM counterparts \citep{2018MNRAS.476.2220L, 2018MNRAS.480.3842O, 2021PhRvD.103j4055W, 2022ApJ...929....9X, 2022MNRAS.509.3772Y}. \citet{2023MNRAS.518.6183M} investigated the detectability of the kilonova and SGRB counterparts of lensed BNS mergers identified by GW detectors; however, they did not employ forward models that directly relate GW parameters to optical light curves. In a different study, \citet{ganguly_more_26} explored the detectability of lensed kilonova in the LSST survey. However, their work is more focused on the serendipitous discovery of the EM counterparts as transients, without depending on the GW detections. 

In this work, we investigate the detectability of the EM counterparts of lensed BNS and NSBH mergers detected during the $\rm O5$ observing scenario (the network configuration is defined in the appendix), with a particular focus on lensed kilonova. First, we simulate a population of lensed BNS and NSBH mergers detectable in the $\rm O5$ observing run. For these detectable events, we employ forward models to generate their corresponding light curves. Specifically, we utilize the BNS generative (BNS-gen) model available in MOSFiT \citep{2018ApJS..236....6G} to generate the kilonova light curves. To generate NSBH kilonova light curves, we employ two models: the NSBH generative (NSBH-gen) model in MOSFiT \citep{Gompertz_2021} and the POSSIS model \citep{2019MNRAS.489.5037B} with a grid tailored for NSBH mergers, subsequently comparing their results. Based on the peak magnitudes of the BNS and NSBH light curves, we estimate the fraction of such events that are jointly detectable by GW detectors during the $\rm O5$ observing run and by LSST \citep{2019ApJ...873..111I}. Furthermore, we explore the GW and lens parameter spaces to identify the configurations that are most favorable for joint detection. 

This paper is organized as follows: Section \ref{methods} describes the simulation of the lensed BNS and NSBH populations and the generation of their respective kilonova light curves. Section \ref{results} presents the results regarding the detectability of lensed BNS and NSBH light curves with LSST, highlighting the GW and lens parameters most favourable to observation. We also show the robustness and generalizability of the detectability index for lensed BNS and NSBH mergers. Finally, Section \ref{conclusion_section} summarizes our findings and provides concluding remarks.

\section{Methods}\label{methods}

We prepare the lensed GW population for BNS and NSBH with the prescription given in \citep{haris2018, Magare_2023}. We consider the lensed binaries to be GW-detectable if any 2 images have SNR greater than 8, in O5 observing scenario. The details of the generated BNS and NSBH population are given in the appendix section.

Once we obtain the data on detectable lensed BNSs and NSBHs, we use this GW data to forward-model the BNS kilonova light curves using the \citep{Nicholl:2021rcr} model and the NSBH kilonova light curve using \citep{2019MNRAS.489.5037B, Gompertz_2021} models.

\subsection{Generating BNS kilonova light curves}
\label{sec-gen_BNS_light_curves}
We use the BNS generative (BNS-gen) model \citep{Nicholl:2021rcr} available in MOSFIT package \citep{2018ApJS..236....6G} to get the BNS kilonova light curves. This is an analytic forward model for
kilonova light curves that is completely specified by the binary
configuration and EoS-dependent properties. The BNS-gen model provides the ejecta masses, velocities and opacities for a given binary configuration, and then calculates the radioactive kilonova light curve for these values using the \citep{2017ApJ...848L..17C, 2017ApJ...851L..21V}. We give a brief description of the model's setup and key assumptions below.

During the merger of two neutron stars, mass is ejected dynamically because of the tidal forces in the orbital plane \citep{2016PhRvD..93l4046S} and also by the shocks at the contact interface between the two neutron stars \citep{2013ApJ...773...78B, PhysRevD.87.024001}. To calculate the dynamical ejecta, BNS-gen uses a fitting formula from \citep{Dietrich_2017} to estimate the ejecta masses from 172 numerical relativity simulations. As encapsulated in the semi-analytic fits of \citep[][see equation~1]{Nicholl:2021rcr}, the properties of the dynamical ejecta are parameterised by the binary mass ratio $q$ alongside the component compactness $C = GM / (Rc^2)$, evaluated for a neutron star of radius $R$. 
$C$ captures the dependence of dynamical ejecta on the EoS. Furthermore, the dynamical ejecta is divided into two components. The red ejecta caused by the tidal forces, having a low electron fraction $(Y_{e} < 0.25)$, are concentrated in the equatorial plane. The blue ejecta caused by the surface shock and neutrino heating at higher latitudes \citep{2016PhRvD..93l4046S, 2020LRR....23....1M} have a higher electron fraction $(Y_{e} < 0.25)$.
The ejecta are termed red and blue because of the higher and lower opacities of the ejecta, respectively. BNS-gen uses simulation results from \citep{2016PhRvD..93l4046S} to estimate the relative mass in the red and blue ejecta. Following the results in \citep{2018ApJ...869..130R}, they fix the grey opacity for each component: $\kappa_{\rm red} = 10$ and $\kappa_{\rm blue} = 0.5$ $\rm cm^2/g$.

Further, additional mass will be ejected after the merger from the surface of the remnant massive neutron star or by the neutrino-driven winds, which come under the category of post-merger ejecta \citep{PhysRevD.95.063016, PhysRevLett.119.231102, 2018ApJ...858...52S, 2018ApJ...869..130R, 10.1093/mnras/sty2932}. The properties of the wind ejecta depend on the lifetime of the remnant, which may be a black hole, hypermassive, supramassive, or stable neutron star \citep{Margalit_2019, 2020LRR....23....1M}. Moreover, the long-lived remnants produce a larger disc mass having a higher $Y_e$. BNS-gen calculates the disc mass as described in 
\citep{2019MNRAS.489L..91C}, where the disk mass $(M_{\rm disk})$ depends on the total mass of the binaries and the threshold mass $(M_{\rm thres})$ for the prompt collapse to a black hole. $M_{\rm thres}$ is proportional to the $M_{\rm TOV}$ and the compactness  \citep{PhysRevLett.111.131101}. A fraction of this disk mass $\epsilon_{\rm disk} \sim 0.1-0.5$ is expected to be ejected by the winds \citep{PhysRevLett.119.231102, 2018ApJ...869..130R}. The typical value of the $\epsilon_{\rm disk}$ is fixed around $\sim 0.2-0.3$ motivated by hydrodynamical and magnetohydrodynamical simulations, which consistently demonstrate that roughly 30\% of the initial disk mass becomes unbound via a combination of viscous heating, magnetic outflows, and neutrino irradiation \citep{2013MNRAS.435..502F, 2023ApJ...942...39F}. 

The velocities of the ejecta in BNS-gen are estimated via \citet{2020LRR....23....1M} using $M_{\rm tot}$, $M_{\rm TOV}$, and $M_{\rm thres}$. Disk ejecta velocity is set at $0.1c$ for stable remnants ($M_{\rm tot} < M_{\rm TOV}$) and $0.03c$ for prompt collapse ($M_{\rm tot} > M_{\rm thres}$), with linear interpolation applied for intermediate masses. For getting the opacity, BNS-gen model uses simulation from  \citep{10.1093/mnras/stx1987} and the $Y_e-\kappa$ relation from \citep{2020MNRAS.496.1369T} to get the purple opacity $(\kappa_{\rm purple})$, which corresponds to the range between $5.5 \ge \kappa_{\rm purple}(\rm cm^2/g) \ge 1.5$. This purple opacity refers to the component of the ejecta material driven by an intermediate electron fraction $(Y_{\rm e}\sim 0.25)$ and a moderate abundance of heavy heavy lanthanide elements \citep{Villar_2017}.  

BNS-gen also considers two additional effects as extensions of this base model. First, the enhancement of the blue ejecta due to magnetically driven winds \citep{2018ApJ...856..101M, Ciolfi_2020}. This is parameterized by $\alpha$ such that $M_{\rm blue,total} = M_{\rm blue,dyn}/\alpha$, with $0< \alpha \leq1$. This mechanism is possible only if the remnant doesn't undergo prompt collapse to a black hole.
The second effect is shock heating of the ejecta by the GRB jet. This effect is included using the formulas from \citep{2018ApJ...855..103P}. The parameter $\theta_{c}$ characterizes this effect, such that the isotropic equivalent luminosity is proportional to the fraction $\theta_{c}^2 /2$ of the total ejecta mass, where $\theta_{c}$ is the half-opening angle of the cocoon.

\begin{table*}[t]
    \centering
    \caption{BNS Kilonova parameters}
    \label{tab:parameters_closed}
    \begin{tabular}{|l|l|} %
        \hline
        \textbf{Parameter} & \textbf{Value} \\ \hline
        Disk fraction ($\epsilon_{\text{disk}}$)    & Uniformly sampled between $0.1-0.3$ \\ \hline
        Enhancement due to magnetically driven winds ($\alpha$)                    & Uniformly sampled between $0.4-0.6$ \\ \hline
        Cocoon half-opening angle ($\cos(\theta_{c})$)    & 0.95 \\ \hline
        Hydrogen column density in host galaxy ($\rm \log N_{H}$) & 19 \\ \hline
        TOV mass of the neutron star ($M_{\rm TOV}$) & 2.3 \\ \hline
        Radius of the neutron star ($R_{\rm NS}$) & 11 \\ \hline
    \end{tabular}
\end{table*}

In Table~\ref{tab:parameters_closed}, we mention the choice of BNS-gen model parameters for generating the BNS kilonova population. The specific choice of these parameters is motivated by the posterior distribution for the agnostic model fit to the GW170817 light curve \citep{2017Sci...358.1556C, 2017Natur.551...75S, Nicholl:2021rcr}. The parameters not mentioned above are set to their fiducial values (see \citep{Nicholl:2021rcr}). 
The additional parameters required to generate the light curves—including chirp mass $(\mathcal{M})$, mass ratio $(q)$, inclination or the observer's viewing angle $(\cos(\theta))$, luminosity distance $(d)$—are sourced from lensed binary parameters from the GW simulation of the O5 observing run which are detectable 

Using the above parameters, we generate the population of kilonova light curves. To generate lensed kilonova light curves, we use the distance modulus formula with magnification factors that scale the kilonova light curves in apparent magnitude. Furthermore, the kilonova images will arrive at different times for the observer due to lensing time delays. In Fig.~\ref{fig:BNS-hist}, we show the lensed kilonova light curves for a quad-lensing configuration, highlighting changes in the apparent magnitude and time delays.

\subsection{Generating NSBH kilonova light curves}
As we have not yet confidently observed a kilonova from an NSBH merger 
the uncertainties inherent in both semi-analytical model parameters and numerical simulations remain substantial. Consequently, we generate NSBH light curves using two complementary methods: (1) the NSBH-genenerative model implemented in MOSFiT \citep{Gompertz_2021}, and (2) the Monte Carlo radiative transfer code POSSIS for NSBH mergers \citep{2019MNRAS.489.5037B, 2021NatAs...5...46A}. We compare the results of these two models to evaluate the observability of NSBH kilonova with LSST.

\subsubsection{MOSFIT model}

We use the NSBH genenerative \citep{Gompertz_2021} (NSBH-gen) model in the MOSFiT package to generate NSBH kilonova light curves. This is a semi-analytical model for NSBH-driven kilonova that predicts the light curves from binary configurations and the neutron star EoS-dependent properties. We give a brief description of the model's setup and key assumptions below.

To produce an EM transient in the case of NSBH mergers, the neutron star must be tidally disrupted by the black hole in the final stages of the inspiral, and some mass must remain outside the event horizon of the black hole \citep{2012PhRvD..86l4007F,PhysRevD.98.081501}. This tidal disruption occurs when the Roche lobe of a neutron star overflows at a distance greater than the innermost stable circular orbit (ISCO) of the black hole. The NSBH-gen model uses the analytical fitting function for the mass of the material that
remains outside the BH event horizon ($M_{\rm ej}$) \citep{PhysRevD.98.081501} and calibrates it with 75 numerical relativity simulations \citep{PhysRevD.79.044024, PhysRevD.83.024005, PhysRevD.85.044015, PhysRevD.87.084006, PhysRevD.90.024026, PhysRevD.84.064018, PhysRevD.92.044028, 2013CQGra..30m5004L, 2018PhRvD..98f3009B}. Using this, the NSBH-gen model parameterizes the ejecta mass in terms of the ISCO radius, the mass ratio of the neutron star to the black hole, the neutron star's compactness, and the neutron star's baryon mass. This implies that the mass of the neutron star ejecta depends primarily on the black hole spin, the binary mass ratio, and the neutron star's equation of state. Furthermore, the dynamical ejecta $(M_{\rm dyn})$ in the NSBH-gen model uses the analytical fitting function of \citep{PhysRevD.101.103002}, with the average velocity of this ejecta modeled by \citep{2016ApJ...825...52K}. The opacity of the dynamical ejecta is modeled as grey opacity, $\kappa = 10 \rm cm^2/g$, following \citep{2013ApJ...775..113T, 2016ApJ...825...52K, 2017Natur.551...80K, 2020MNRAS.496.1369T}. 

Post-merger hydrodynamic simulations demonstrate that the energy released from nuclear recombination and viscous heating ejects part of the post-merger disc as neutron-rich outflows \citep{2013MNRAS.435..502F, 2015MNRAS.446..750F, 2015MNRAS.448..541J, 2020MNRAS.497.3221F, PhysRevD.101.083029}. In NSBH-gen model, the disc mass is calculated as $M_{\rm disc} = M_{\rm ej} - M_{\rm dyn}$ and the mass of material driven from the disc by thermal pressure $(M_{\rm th})$ is given by $M_{\rm th} = M_{\rm disc}\left(\xi_{1} + \frac{\xi_{2}-\xi_{2}}{1+e^{1.5/(q-3)}}\right)$, which uses the inferences from \citep{2020MNRAS.497.3221F, 2021ApJ...922..269R}. This ejected material is assumed to have an averaged velocity $v_{\rm therm} = 0.034c$ 
\citep{2020MNRAS.497.3221F} and grey opacity of $\kappa \sim 5 \rm cm^2/g$ \citep{2020MNRAS.496.1369T}. Furthermore, three-dimensional general relativistic magnetohydrodynamic simulations show that the presence of another outflow is mediated by magnetohydrodynamic winds \citep{PhysRevLett.119.231102, 2019MNRAS.490.4811C, 10.1093/mnras/sty2932}. In the NSBH-gen model, this magnetically driven wind component is accounted for by the parameter $f_{\rm mag}$ as $M_{\rm mag} = f_{\rm mag}M_{\rm th}$ and $v_{\rm mag} = f_{\rm mag}0.22c$. Physically, $f_{\rm mag}$ captures the geometry of the post-merger magnetic fields: $f_{\rm mag} = 1$ corresponds to the full poloidal magnetic field and hence more ejecta mass and higher velocities \citep{ 10.1093/mnras/sty2932}, and lower values correspond to more toroidal geometries. This magnetically driven ejecta has $Y_{\rm e}\sim 0.1$ which corresponds to $\kappa\sim 10 \rm cm^{2}/g$. 

The geometry of the NSBH-gen model is similar to that of the BNS-gen model \citep{Nicholl:2021rcr}. Following the formalism of \citep{Darbha_2020}, the emitting regions are modeled by scaling the luminosity of each component to the projected area of the caps as seen by an observer at a viewing angle $\theta_{\text{\rm obs}}$. Further, the magnetic winds are restricted to the polar regions with $\theta_{\rm mag} > 45^{\rm o}$, the thermal winds around $80^{\rm o}$, and the dynamical ejecta around $\pm 10^{\rm o}$ from the equator. The NSBH-gen model assumes that the different emitting regions do not interact. Further, this model is based on simulations in which the black hole spin axis is aligned with the orbital spin axis, so there is no spin precession. 

\begin{table*}[t]
    \centering
    \caption{NSBH Kilonova parameters}
    \label{tab:nsbh_parameters_closed}
    \begin{tabular}{|l|l|} %
        \hline
        \textbf{Parameter} & \textbf{Value} \\ \hline
        Opening angle of wind-dynamical ejecta ($\cos(\theta_{\rm wind})$)    & Uniformly sampled between $0.0,0.342$ \\ \hline
        Opening angle of magnetic winds ($\cos(\theta_{\rm mag})$)                    & Uniformly sampled between $0.1,0.5$ \\ \hline
        Orbit-aligned component of black hole spin ($\chi_{\rm BH}$) 
                      & Uniformly sampled between $0.3,1.0$ \\ \hline
        Magnetic wind fraction ($f_{\rm mag}$)    & Uniformly sampled between $0.1,1$ \\ \hline
        Hydrogen column density in host galaxy ($\rm \log N_{H}$) & 19 \\ \hline
        TOV mass of the neutron star ($M_{\rm TOV}$) & 2.3 \\ \hline
        Radius of the neutron star ($R_{\rm NS}$) & 11 \\ \hline
    \end{tabular}
\end{table*}

In Table~\ref{tab:nsbh_parameters_closed}, we list the NSBH-gen model parameters used to generate the NSBH kilonova population. The parameters not mentioned above are set to their fiducial value, see \citep{2018ApJS..236....6G, Gompertz_2021}. 
The additional parameters required to generate the light curves --- including the mass of the black hole $(m_{\rm 1})$, mass of the neutron star $(m_{\rm 2})$, inclination $(\cos(\theta))$, and the luminosity distance $(d)$--- are sourced from the lensed binary parameters from the GW simulation fo the O5 observing run which are detectable. 

Further, as we did for generating the lensed BNS kilonova light curves, we use magnification factors and lensing time delays to obtain the lensed NSBH kilonova light curves. 

\subsubsection{ POSSIS model}
We use the Monte Carlo radiative transfer code POSSIS \citep{2019MNRAS.489.5037B}, with a grid tailored for NSBH mergers \citep{2021NatAs...5...46A}, to generate NSBH light curves. We give a brief description of the simulation's setup and highlight the assumptions below.


POSSIS (Polarization Spectral Synthesis in Supernovae) is a three-dimensional Monte Carlo radiative transfer code designed to simulate polarization and synthesize multi-wavelength light curves for supernova and kilonova. In this work, we employ the precomputed NSBH kilonova simulation grid introduced by \citet{2021NatAs...5...46A}. This model grid discretizes the synthetic observables on a two-component ejecta morphology:
\begin{enumerate}
    \item \textbf{Dynamical ejecta ($M_{\rm ej,dyn}$):} A cold, highly neutron-rich, and lanthanide-rich tidal component concentrated in the equatorial plane within a half-opening angle $\pm\phi$.
    \item \textbf{Wind ejecta ($M_{\rm ej,wd}$):} A post-merger secular accretion-disk wind modeled as a spherical outflow, characterized by intermediate electron fractions ($Y_e$) corresponding to matter with opacity intermediate between lanthanide-poor and lanthanide-rich compositions.
\end{enumerate}
Across this grid, simulated spectra and broadband light curves are evaluated over a discrete parameter space spanning the ejecta masses ($M_{\rm ej,dyn}$ and $M_{\rm ej,wd}$), the tidal tail half-opening angle $\phi$, and the observer viewing angle $\theta_{\rm obs}$.

NSBH grid for this simulation provides the spectral energy distributions (SEDs) for $81$ models with varying ejecta masses and a fixed $\phi = 30^{\rm o}$. The ejecta masses for the $2$ components spans: $M_{\rm ej,dyn}$, $M_{\rm ej, wd}\in [0.01, 0.09] M_{\odot}$ with the step size of $0.01 M_{\odot}$. For each of these models, $11$ different viewing angles 
are utilized, which are linearly spaced from pole $\cos\theta_{\rm obs} = 1$ to equator $\cos\theta_{\rm obs} = 0$. Further, for each model, the flux is given in units of $\rm erg$ $\rm s^{-1}$ $\rm cm^2$ $\rm \AA^{-1}$ at a distance of $10$pc. The time step is linearly spaced in $100$ intervals from $0.1$ to $20$ days, and the wavelength is also linearly spaced in the intervals of $500$~$\rm \AA$ from $100$ to $99900$~$\rm \AA$. 

To obtain the light curves for the simulated NSBH population, the dynamical ejecta mass ($M_{\rm ej,dyn}$) is set equal to the dynamical mass calculated by the NSBH-gen model. The wind ejecta mass, $M_{\rm ej, wd}$, is defined as the sum of the thermal ($M_{\rm th}$) and magnetic ($M_{\rm mag}$) components from the same model. While the dynamical velocity is adopted directly from the NSBH-gen output, the wind velocity is calculated as the mass-weighted average of the thermal and magnetic velocities, i.e., $V_{\rm wind} = \frac{M_{\rm th}V_{\rm th} + M_{\rm mag}V_{\rm mag}}{M_{\rm th} + M_{\rm mag}}$. 

To ensure physical consistency and avoid extrapolation, the analysis is restricted to the NSBH population with ejecta masses falling within the existing simulation grid. This grid spans $[0.01, 0.09] M_{\odot}$ for both $M_{\rm ej,dyn}$ and $M_{\rm ej, wd}$. For each merger event, the calculated ejecta parameters are mapped to the nearest grid point to generate the corresponding light curve. Finally, the resulting light curves are K-corrected for the LSST \textit{ugrizy} bands using the 81 Bu2019nsbh POSSIS SED files and the \texttt{sncosmo} package. Furthermore, we use the magnification factors and lensing time delays to generate the lensed NSBH kilonova light curves. 

\subsection{Detectability index for lensed BNS and NSBH population}
We start with the GW-detectable lensed BNS and NSBH mergers for the O5 observing run and employ forward models to generate their corresponding light curves. We then use LSST's exposure limits to determine the population that is also EM detectable (i.e., jointly GW$+$EM detectable). We propose a detectability index ($\mathrm{DI}$), which takes the lensed GW parameters as the features and returns a value that determines how likely the GW-detectable lensed BNS or NSBH merger is also EM-detectable. The primary objective is to formulate a robust, physically interpretable detectability metric capable of separating the lensed GW$+$EM detectable population 
against the lensed GW detectable population.  

\begin{figure*}
\includegraphics[scale = 0.44]
{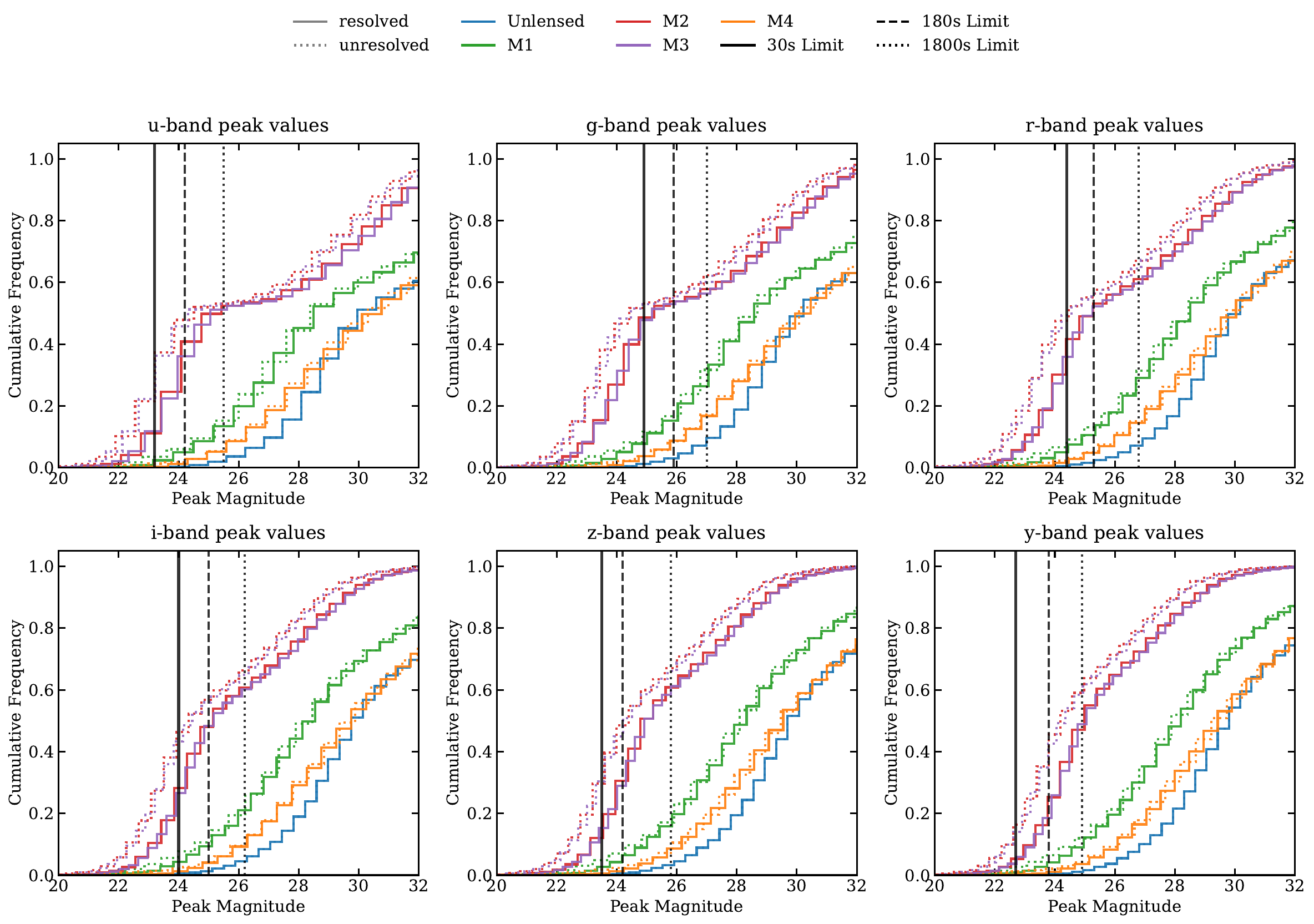}
\caption{Distribution of the peaks of lensed BNS light curves (rendered as cumulative histograms) for different images is plotted for different LSST filters. The images $\rm M1, M2, M3, M4$ are in the order of their arrival times for quads. The doubles are included in the $\rm M1, M2$. The distribution of peaks for the unlensed BNS light curves is also plotted. The solid and dashed lines correspond to the cases of lensed images that are resolved and unresolved, respectively. The criterion we use for unresolved: if any two images have angular separations smaller than $0.8''$ and time delays smaller than $1$ day, the flux is added. The vertical lines correspond to the various LSST exposure limits. The fraction of events to the left of the exposure limit thresholds are considered detectable.}
\label{fig:BNS-hist}
\end{figure*}

We approach this by framing it as a binary classification problem. We treat purely GW-detectable lensed BNS and NSBH mergers as one class, and the jointly GW$+$EM detectable mergers as the other. We then define the detectability index for BNS and NSBH, motivated by logistic regression\citep{Cox1958BinarySequences, Hosmer2013Applied}, as follows 
\begin{equation}
    \begin{split}
        \mathrm{DI_{BNS}} =& \ c_{1}\log(\mu) + c_{2}\log(M) + c_{3}\log(q) \\
        &+ c_{4}\log(|\cos \iota|) + c_{5}\log(D_{L})
    \end{split}
\end{equation}
\begin{equation}
    \begin{split}
        \mathrm{DI_{NSBH}} =& \ d_{1}\log(\mu) + d_{2}\log(M) + d_{3}\log(q) \\
        &+ d_{4}\log(|\cos \iota|) + d_{5}\log(D_{L}) + d_{6}\log(\chi)
    \end{split}
\end{equation}
where $M$ is the total mass of the binaries, $q$ is the mass ratio, $\iota$ is the inclination angle, $D_L$ is the luminosity distance to the binaries, $\chi$ is the spin of the black holes in NSBH case, and $\mu$ is the magnification of the brightest image. 

This metric is a multivariate linear combination of intrinsic and extrinsic physical parameters, evaluated on a log-parameter space, and modeled as a logistic regression classifier \citep{Hosmer2013Applied}, implemented via \texttt{scikit-learn} \citep{scikit-learn}.  
By extracting the fitted coefficients by training the model, we obtain the optimal weights for each physical parameter \citep{Hastie2009Elements, Bishop2006Pattern}. The parameters with positive weights improves detectability, while those with negative weights suppress detectability. This process yields a data driven, empirical metric that maps to the physical dynamics of the system. 

\section{Results and Discussion}\label{results}


We present the results regarding the joint detectability of lensed BNS and NSBH populations with GW in the O5 observing run and in EM with LSST. 
Furthermore, we highlight the EM and GW properties characterizing this detectable lensed population.

The detectability of a multi-messenger (GW+EM) lensed event is defined by the following criteria:
\begin{itemize}
    \item \textbf{Lensed GW threshold:} A signal-to-noise ratio ($\text{SNR}$) greater than $8$) for at least two lensed images.
    \item \textbf{Lensed EM Counterpart:} The peak AB apparent magnitude of one of the images of the light curve must be brighter than the LSST exposure limit threshold in any one of the observing bands.
\end{itemize}

\subsection{BNS Population Statistics}


Once we have generated all lensed BNS light curves as described in Section~\ref{sec-gen_BNS_light_curves}, we analyze their statistics. We have a total population of $3125$ lensed BNS light curves that satisfy the lensed GW detection threshold for the O5 observing run. In Fig.~\ref{fig:BNS-hist}, we show the cumulative histogram for this lensed BNS population (\textit{solid lines}) in $6$ LSST bands ($u, g, r, i, z,$ and $y$), along with the exposure limits for each band. The single exposure limit (i.e., 30 seconds of exposure time) refers to the limiting magnitude at which the faintest signal can be detected with $5\sigma$ above the background noise. From Fig.~\ref{fig:BNS-hist}, we can see that only a small fraction of the unlensed population ($\leq 1\%$) lies above the single exposure limit of LSST. However, due to lensing magnification, this fraction is significantly enhanced. 

\begin{figure*}
\centering
\includegraphics[scale = 0.38]{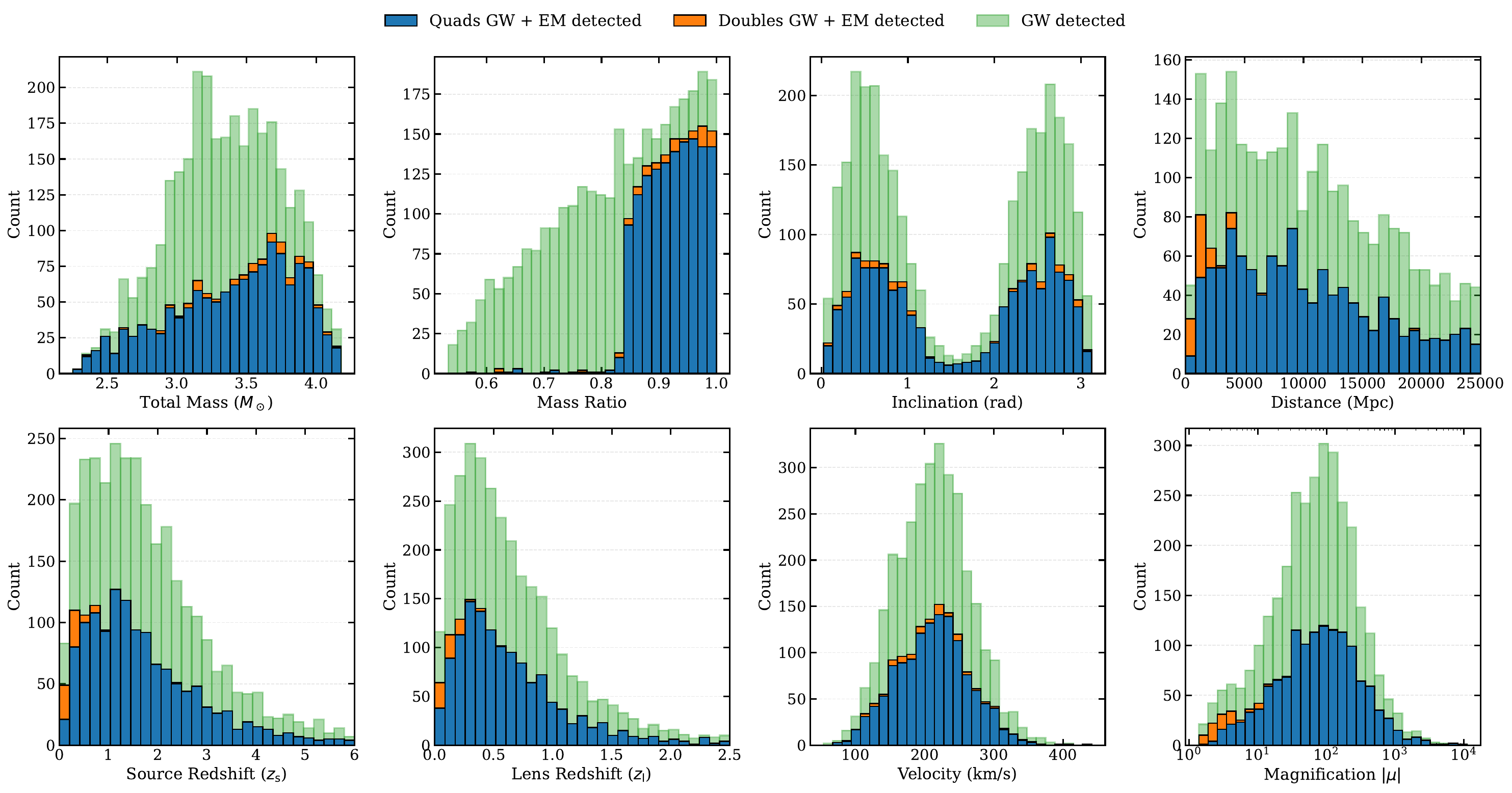}
\caption{This distribution of GW and lens system parameters for the lensed GW detected population (\textit{light green}) and the lensed GW + EM detected population (\textit{dark green} corresponds to the quads and \textit{orange} corresponds to the doubles). The EM detection criterion of g-band magnitude $\le$ 24.9 is used, which corresponds to the 30s exposure limit.  }
\label{fig:BNS-GW-params}
\end{figure*}

\begin{figure*}
\centering
\includegraphics[scale = 0.38]{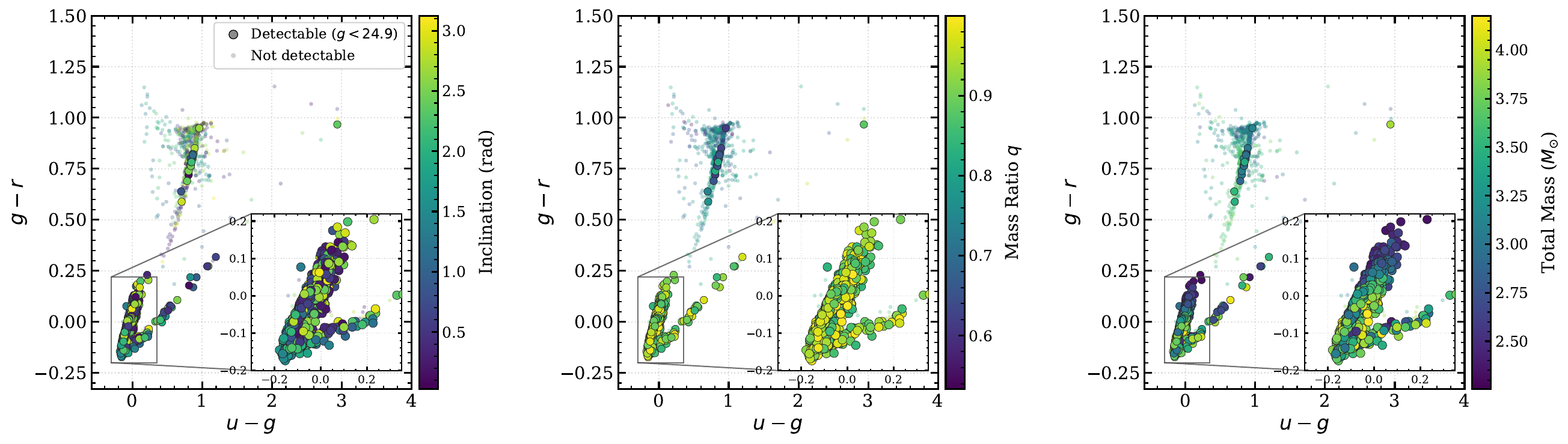}
\caption{The color-color (u-g \textit{vs} g-r) plot of GW parameters: Inclination, Mass ratio, Total mass is plotted for the GW detected systems. The color bar shows the values of the corresponding GW parameter in each panel. The highlighted dots correspond to the GW parameters of lensed GW+EM detectable systems.}
\label{Fig:BNS-color-GW-all}
\end{figure*}

\begin{figure*}
\centering
\includegraphics[scale = 0.38]{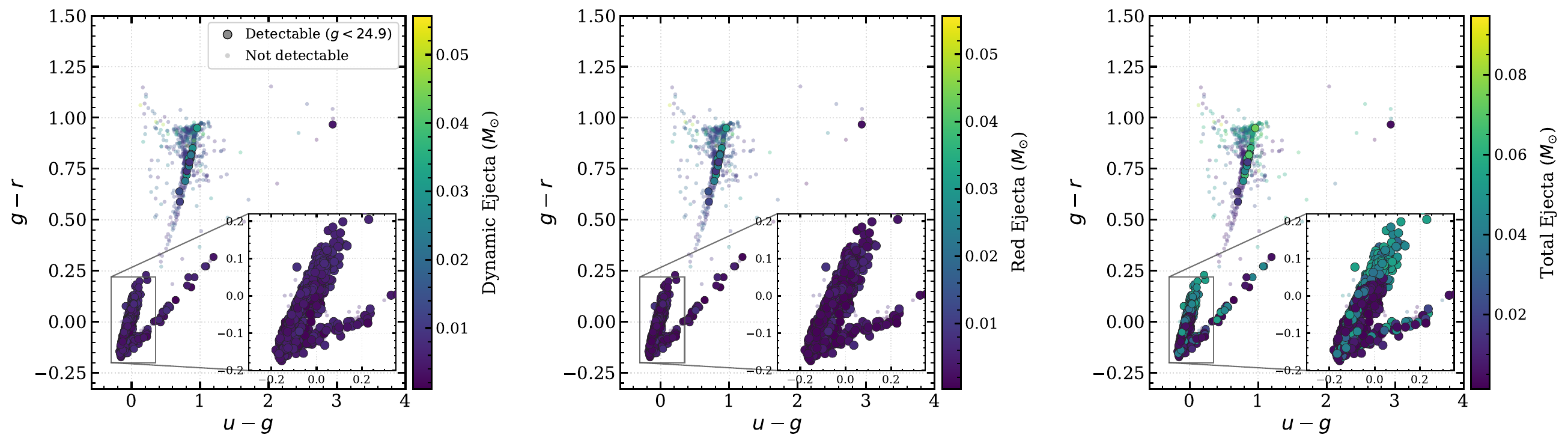}
\caption{The color-color (u-g \textit{vs} g-r) plot of ejecta parameters: Dynamical, Red, Total is plotted for the GW detected systems. The color bar shows the values of the corresponding GW parameter in each panel. The highlighted dots correspond to the GW parameters of lensed GW+EM detectable systems.}
\label{Fig:BNS-color-GW-ejecta}
\end{figure*}

As the lensed images of the light curves are labeled in order of arrival time, the $2\rm nd$ (\textit{red}) and the $3\rm rd$(\textit{purple}) image, which have the highest magnifications, show significant flux increase across all bands. When we compare the fraction of the population above the single exposure limit across the band, we see that the $g$, $r$, and $i$-bands show a higher detectable fraction than the other $3$ bands. Out of which the $g$-band shows the highest detectable fraction of $51\%$, followed by $40\%$ in the $r$-band, and then $30\%$ in the $i$-band. We also note that the detectable fractions reach $60\%$ at a much longer exposure time of $1800$ seconds, although this is not practical for a realistic observational survey \citep{ganguly_more_26} 
The detectability of other lensed images ($1\rm st$ and $4\rm th$) appears unlikely at these exposure limits, and a targeted, deeper observation is required. 



Further, we implement an unresolved images criterion on our population of lensed BNS light curves, using the criterion that if any two images have angular separations smaller than $0.8''$ and time delays smaller than $1$ day, we add their fluxes. In Fig.~\ref{fig:BNS-hist}, we show the cumulative histogram for the lensed BNS population (\textit{dashed lines}) after accounting for the unresolved images. We find that accounting for the unresolved image criterion yields a larger number of lensed kilonova peaks above the single exposure limit of LSST. Further, the $g$-band detectable fraction has increased to $58\%$ after the unresolved image criterion, compared to $51\%$ in the resolved scenario. We observe a similar increase in the detectable fraction after taking into account the unresolvability of lensed images across all LSST bands. 

We compare the GW parameters of lensed BNS systems whose $g$-band light curves are detectable by LSST at the single-exposure limit, accounting for the unresolved image criterion.   In Fig.~\ref{fig:BNS-GW-params}, we present the distributions of total mass, mass ratio ($q$), inclination ($\iota$), luminosity distance ($D_L$), source redshift ($z_s$), lens redshift ($z_l$), velocity dispersion ($\sigma$), and magnification ($\mu$). These distributions are shown for two distinct groups: the population detected only by GW and the population detectable by both GW and EM. Fig.~\ref{fig:BNS-GW-params} highlights that quad-image systems are significantly more numerous than double-image systems within the multi-messenger population. 
The redshift and magnification distributions reveal that detectable lensed BNS events, which are quads, typically originate from a broad source redshift distribution peaking at $z_s \sim 1$, with lenses peaking at $z_l \sim 0.5$ and magnifications on the order of $\mu \sim 100$. While detectable lensed BNS events, which are doubles, originate at lower redshifts and have lower magnifications. While most GW+EM parameters track the underlying GW-only distributions, the mass ratio exhibits a distinct divergence. We find a sharp cut-off at $q \sim 0.82$, indicating that EM detection in the $g$-band strongly favors symmetric mass binaries. This preference arises because symmetric binaries undergo head-on merger 
that produce significant shock-heated ejecta with higher electron fractions ($Y_e$). This material generates a ``blue" kilonova that is intrinsically brighter at shorter wavelengths, thereby boosting the detectability of symmetric systems within the LSST $g$-band.

In Fig.~\ref{Fig:BNS-color-GW-all}, we present the color--color diagram for the GW-detectable lensed BNS population, which exhibits a characteristic bifurcated (split) distribution. This distinct split in the population is driven by the combined effects of three physical parameters: inclination ($\iota$), mass ratio ($q$), and total mass ($M_{\text{tot}}$). 

Higher inclinations ($\iota \sim 1.5$--$3.0$\, rad) correspond to observations through the lanthanide-rich equatorial ejecta; this material effectively obscures blue emission, resulting in a redder observed kilonova. Conversely, lower inclinations ($\iota \sim 0$--$0.5$\, rad) align with the polar axis, where the blue kilonova component remains visible. Furthermore, lower mass ratios (increased asymmetry) produce redder colors due to enhanced tidal ejecta, which is inherently lanthanide-rich. While the total mass has a comparatively weaker impact on the population split, it determines the positioning of events within these clusters. Specifically, higher total masses increase the likelihood of prompt collapse, thereby reducing the total ejecta mass. This reduction in ejecta shifts the points toward the blue region of the diagram (down and to the left), as a lower ejecta opacity allows the hot, early-time blue peak to be more easily observed.

In Fig.~\ref{Fig:BNS-color-GW-ejecta}, we present color-color diagrams for the GW-detectable lensed BNS population, with the color scales representing the mass of dynamical, purple, red, and total ejecta components. A comparison between Fig.~\ref{Fig:BNS-color-GW-all} and Fig.~\ref{Fig:BNS-color-GW-ejecta} clearly demonstrates that lower mass ratios ($q$) tend to produce higher red ejecta masses, whereas more symmetric binaries ($q \approx 1$) primarily emit bluer ejecta. A similar correlation is observed between the mass ratio and the dynamical ejecta mass. Furthermore, the total mass distribution in Fig.~\ref{Fig:BNS-color-GW-all} correlates strongly with the total ejecta mass shown in Fig.~\ref{Fig:BNS-color-GW-ejecta}; specifically, binaries with higher total masses produce lower total ejecta masses, and vice versa. Furthermore, the lensed BNS population detectable via both GW and EM (in the $g$-band) is highlighted by the larger dots. Our results indicate that the EM-detectable population strongly favors symmetric mass ratios and shows a slight preference for higher total masses. This is due to the fact that symmetric mass binaries undergo head-on collisions, producing brighter, blue, shock-heated ejecta that is more easily detected at shorter wavelengths, such as the LSST $g$-band. This physical interpretation is supported by Fig.~\ref{Fig:BNS-color-GW-ejecta}, which shows that the multi-messenger population is characterized by binary mergers with lower dynamical and red ejecta masses, as well as a slight preference for lower total ejecta masses.


\begin{figure*}
\centering
\includegraphics[scale = 0.44]{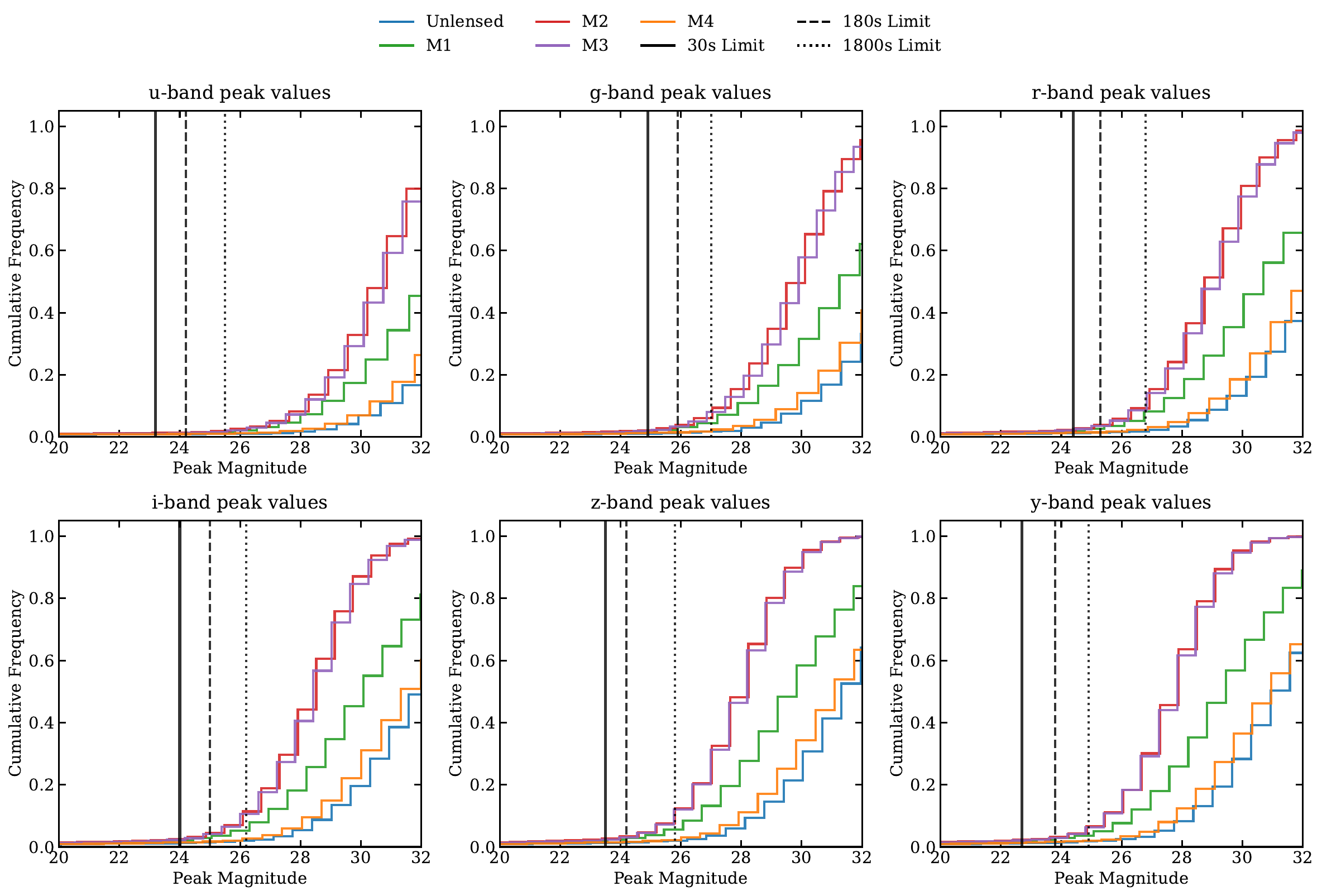}
\caption{Distribution of the peaks of lensed NSBH light curves using \textbf{NSBH-gen model} (rendered as cumulative histograms) for different images is plotted for different LSST filters. The images $\rm M1, M2, M3, M4$ are in the order of their arrival times for quads. The doubles are included in the $\rm M1, M2$. The distribution of peaks for the unlensed NSBH light curves is also plotted. The vertical lines correspond to the various LSST exposure limits. The fraction of events to the left of the exposure limit thresholds are considered detectable.}
\label{Fig:NSBH-hist}
\end{figure*}


Given the low expected detection rates, discovering a lensed BNS with joint GW and EM counterparts remains an unlikely prospect in upcoming observing runs. However, should an event be detected, it will almost certainly be an outlier featuring extreme magnification ($\mu \sim \mathcal{O}(100)$) in a quad-image configuration, biased toward higher total binary masses and nearly symmetric mass ratios.


\subsection{NSBH Population Statistics}
From an overall population of 31,239 lensed NSBH events detectable via gravitational waves during the O5 observing run, we use the NSBH-gen model to synthesize counterpart kilonova light curves. Of these, 9,019 events produce observable light curves. The presence of a kilonova in an NSBH merger depends upon whether the neutron star undergoes tidal disruption prior to being swallowed by the black hole. This disruption depends primarily on three factors: the mass ratio of the system ($q$, where $0 < q < 1$), the dimensionless spin of the black hole ($\chi_{\rm bh}$), and the equation of state (EoS) of the neutron star. In our population, we fix the EoS by defining the maximum TOV mass ($M_{\rm TOV}$) and the neutron star radius ($R_{\rm NS}$); consequently, tidal disruption becomes a function of $q$ and $\chi_{\rm bh}$ alone. Generally, a kilonova is expected in systems characterized by a high mass ratio i.e. more symmetric systems 
and a rapidly spinning black hole.

\begin{figure*}
\centering
\includegraphics[scale = 0.35]{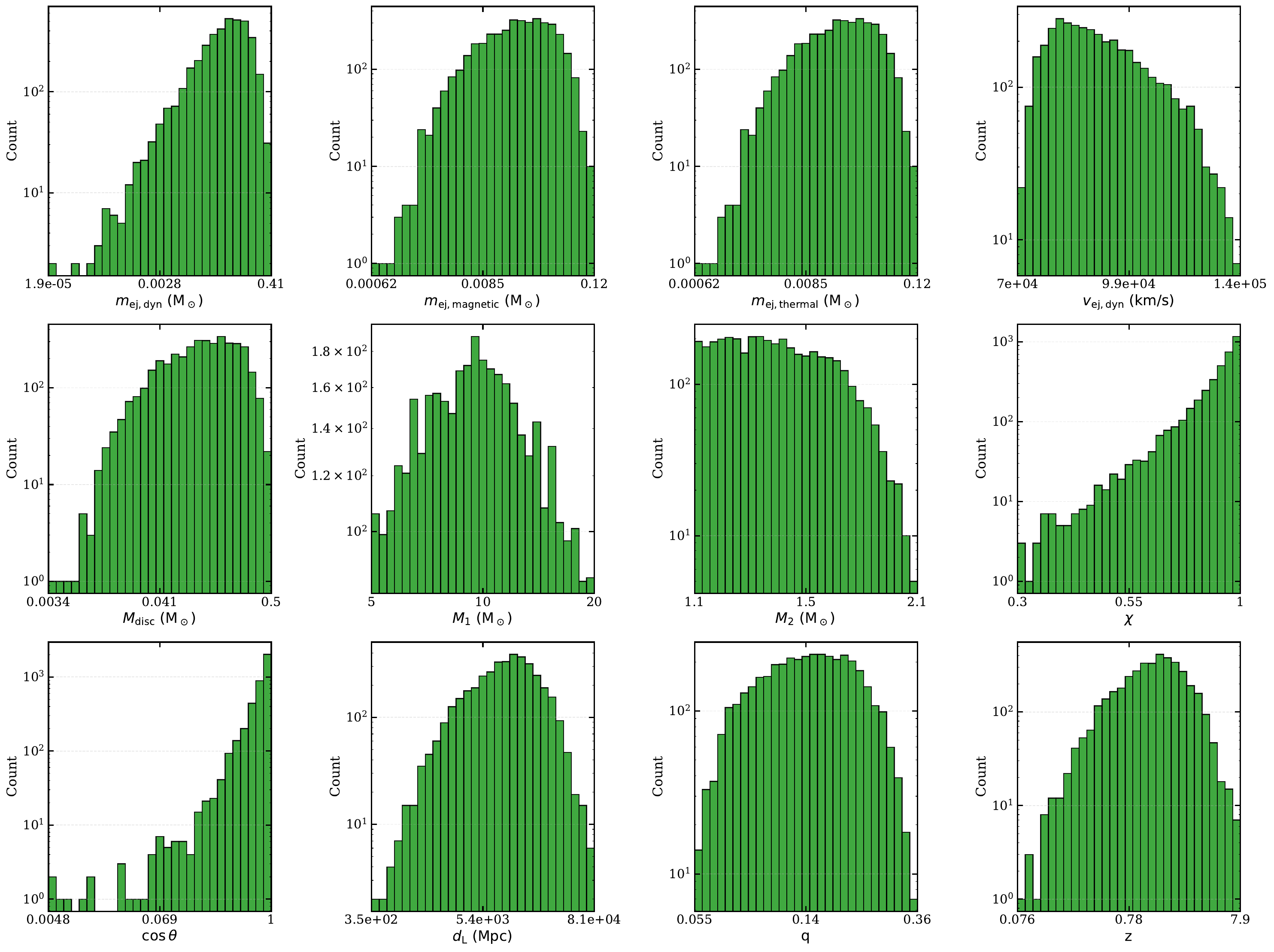}
\caption{Distribution of the GW, ejecta, and lens parameters of lensed NSBH systems, which are GW detectable and non-zero ejecta mass, is plotted. }
\label{Fig:NSBH-intermediate_params}
\end{figure*}

In Fig.~\ref{Fig:NSBH-hist}, we show the distribution of the peak light curve magnitudes and the cumulative histogram for this lensed NSBH population in $6$ LSST bands ($u, g, r, i, z,$ and $y$), along with the limiting magnitudes for each band for different exposure times. 
The Fig.~\ref{Fig:NSBH-hist} illustrates that even after lensing, at most, $2.3\%$ of the NSBH population that produce kilonova light curve is detectable above the single exposure limit in LSST, and $10.4\%$ of the population has a peak brighter than $1800$ second exposure, with slight more fraction in the $r-$band. This is significantly lower than the lensed BNS population. 

\begin{figure*}
\centering
\includegraphics[scale = 0.38]{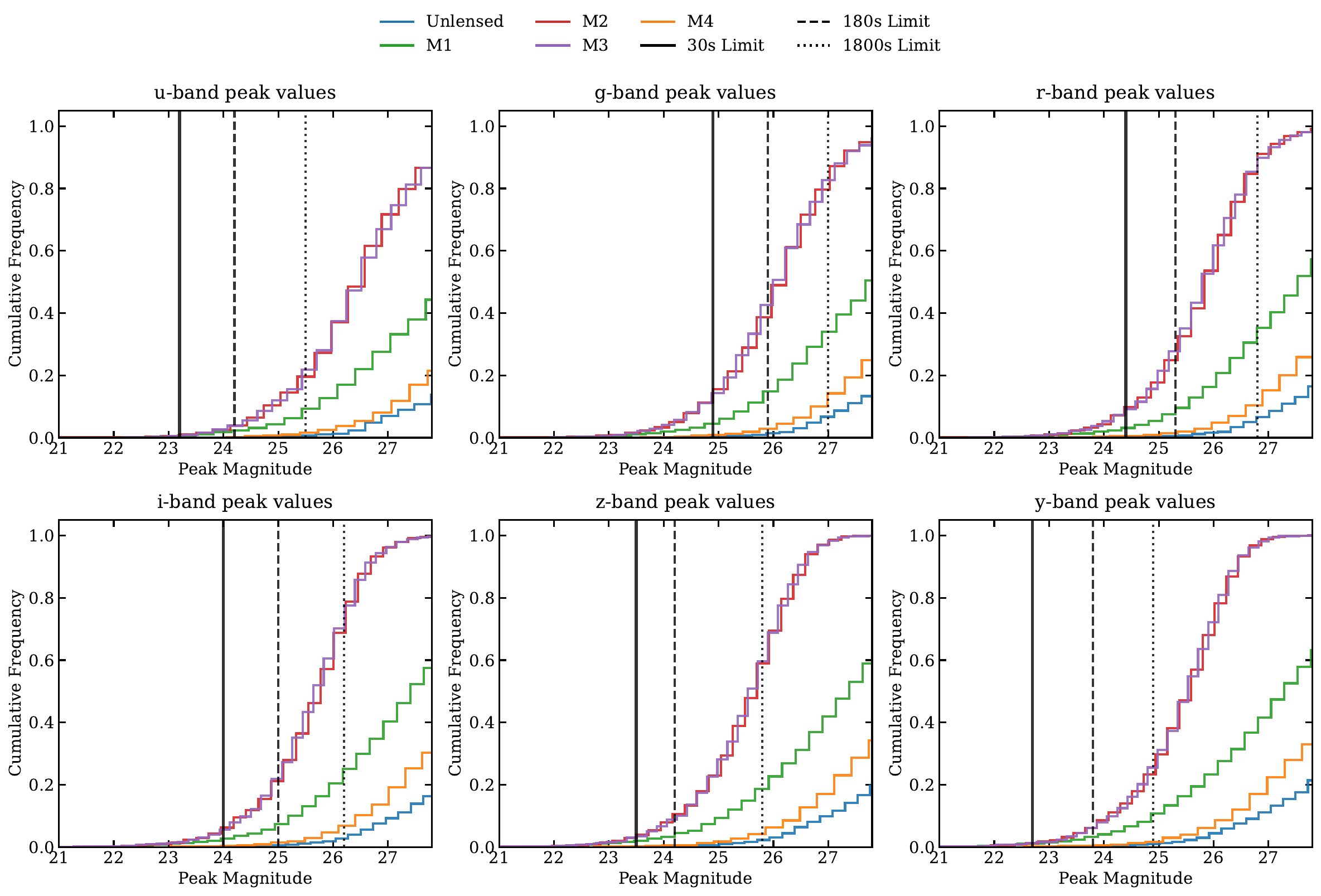}
\caption{Distribution of the peaks of lensed NSBH light curves using \textbf{POSSIS model} (rendered as cumulative histograms) for different images is plotted for different LSST filters. The images $\rm M1, M2, M3, M4$ are in the order of their arrival times for quads. The doubles are included in the $\rm M1, M2$. The distribution of peaks for the unlensed NSBH light curves is also plotted. The vertical lines correspond to the various LSST exposure limits. The fraction of events to the left of the exposure limit thresholds are considered detectable.}
\label{Fig:NSBH-hist-NMMA}
\end{figure*}

For NSBH mergers detectable in the O5 observing run with kilonova emissions—we illustrate in Fig.~\ref{Fig:NSBH-intermediate_params} the distribution of dynamical ejecta ($M_{\rm dyn}$), magnetic ejecta mass ($M_{\rm mag}$), thermal ejecta mass ($M_{\rm ther}$), disc mass ($M_{\rm disc}$), velocity of dynamical ejecta ($v_{\rm dyn}$), mass of the black hole ($M_{1}$), mass of neutron star ($M_2$), mass ratio ($q$), cosine of the inclination ($\cos\iota$), dimensionless spin of the black hole ($\chi_{\rm bh}$), luminosity distance ($d_{L}$) and source redshift of the merger ($z_{s}$). We find that the GW detectable NSBH mergers, which have a higher $\chi_{\rm bh}$, tend to produce a kilonova light curve. As already discussed, this is because the rapidly spinning black hole tidally disrupts the neutron star before it is swallowed. Similarly, the kilonova-producing population exhibits a strong preference for face-on orientations ($\cos \iota \to 1$). This can be attributed to the GW detectors being more sensitive to the face-on orientations. Furthermore, unlike BNS mergers, the intrinsic luminosity of NSBH kilonova peaks towards the equatorial plane. As the detected population is dominated by face-on configurations, we expect to observe lower fluxes in the kilonova light curves than in those with equatorial orientations. We can also see a weaker trend that higher values of  $M_{\rm dyn}$, $M_{\rm mag}$, $M_{\rm ther}$, and $M_{\rm disc}$ are more favored. 

We determine the light curves using the Bu2019nsbh model, which takes $M_{\rm dyn}$, $M_{\rm mag}$, $M_{\rm ther}$, and the corresponding velocities as inputs within the parameter space of Bu2019nsbh model. In Fig.~\ref{Fig:NSBH-hist-NMMA}, we show the cumulative histogram for this lensed NSBH population in 6 LSST bands, along with the exposure limits for each band. Again, the $2$nd and $3$rd images show a significant increase in flux across all bands due to lensing magnification. For this distribution, when we compare the fraction of the population above the single exposure limit across the band, we find that the $g-$ and $i-$ bands show a higher detectable fraction than other bands. Out of which $g-$band shows the highest detectable fractions of $\sim 15\%$ and $\sim 10\%$ in $r$-band. We also note that the detectable fractions reach $\sim 85\%$ at a much longer exposure time of $1800$ seconds. 


Comparing the cumulative distribution of peak light curve magnitudes generated by the NSBH-gen model (Fig.~\ref{Fig:NSBH-hist}) and the `Bu2019nsbh' model (Fig.~\ref{Fig:NSBH-hist-NMMA}) reveals stark differences. For the single-exposure magnitude limit, the detectable fraction of kilonova from the NSBH-gen model is drastically lower ($\sim 2.3\%$ in the $r$-band) than those from the `Bu2019nsbh' model ($\sim 15\%$ in $g$-band). This discrepancy in peak brightness and timing stems primarily from how each model treats computational frameworks, ejecta geometry, and opacities.

The NSBH-gen model relies on a semi-analytical framework featuring three non-interacting, axially symmetric conical regions: dynamical ejecta, a thermally driven wind, and a magnetically driven wind. Crucially, this model lacks neutrino irradiation from the remnant neutron star, which keeps the electron fraction ($Y_{e}$) of the ejecta very low. Because of this, it assigns a high, fixed grey opacity ($\kappa=10\rm cm^{2}g^{-1}$) to its massive, lanthanide-rich components rather than using detailed radiative transfer or atomic data. This high-opacity material heavily suppresses optical emission, resulting in early optical light curves that are both exceedingly faint and significantly delayed in reaching their peak.

In contrast, the `Bu2019nsbh' model utilizes a two-component geometry consisting of dynamical ejecta and a spherical post-merger wind. Unlike the fixed opacities of the generative model, `Bu2019nsbh' natively handles time- and wavelength-dependent opacities across different bands. Its post-merger wind features an intermediate composition between lanthanide-poor and lanthanide-rich material, yielding a lower overall opacity. This allows a substantial amount of early optical light to escape, producing a kilonova light curve that is much brighter and peaks much earlier than its NSBH-gen counterpart.

\begin{figure*}
\label{Fig:BNS_NSBH_DI}
\centering
\includegraphics[scale = 0.38]{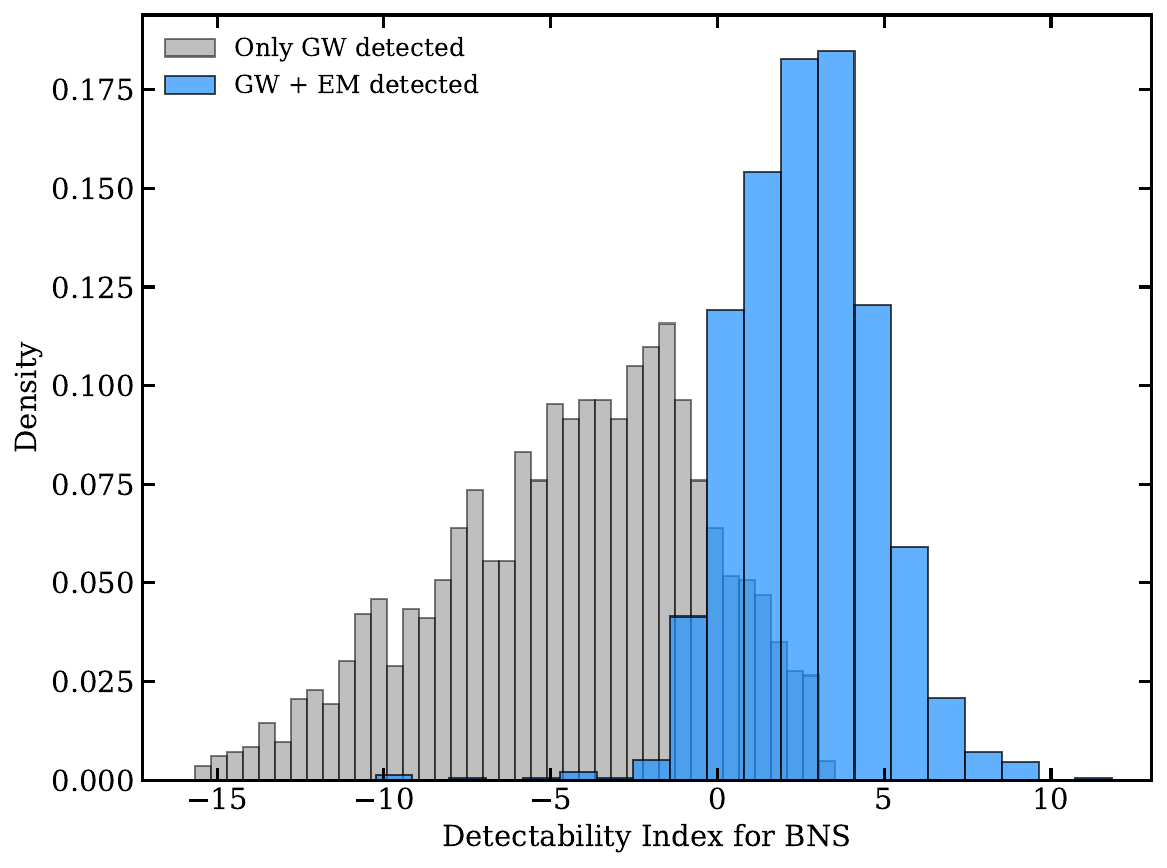}
\includegraphics[scale = 0.38]{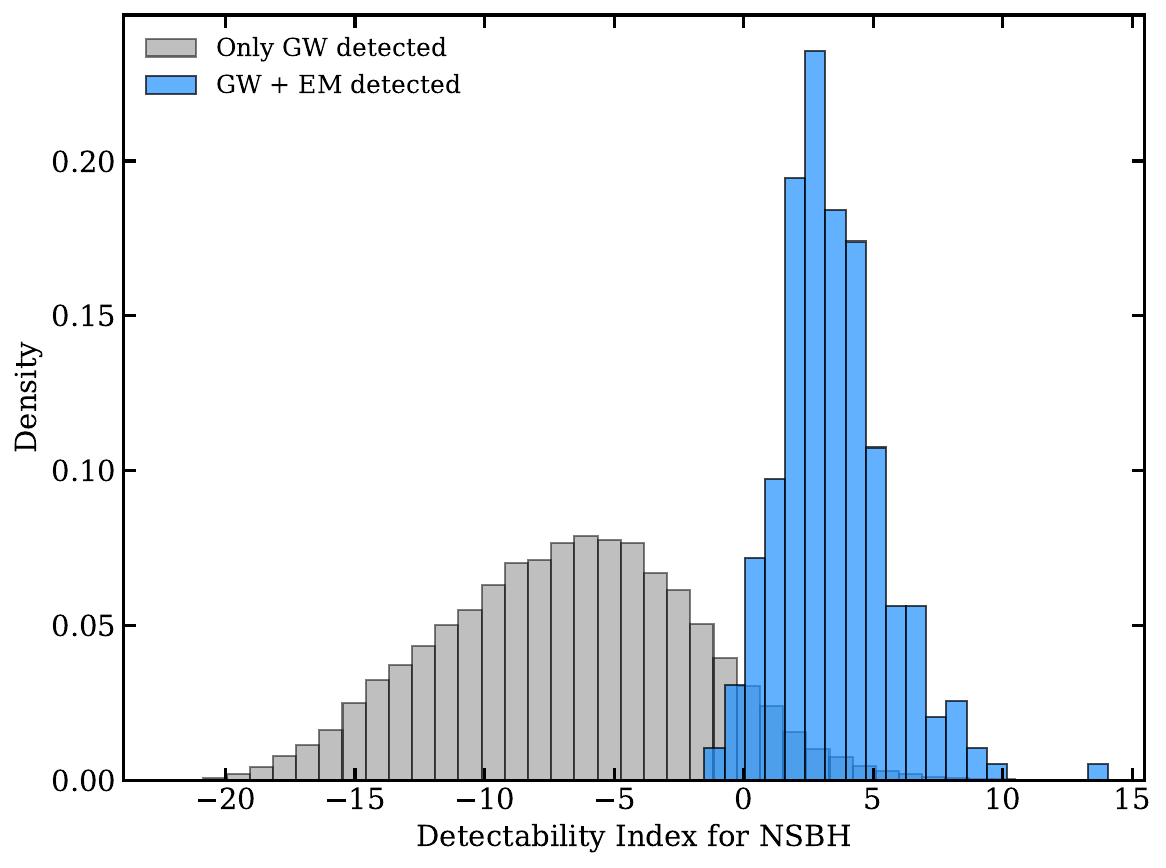}
\caption{The distribution of the Detectability Index for the lensed GW detectable population and the lensed GW + EM detectable population is plotted for BNS (\textit{left}) and NSBH (\textit{right}) systems.}
\label{Fig:BNS_NSBH_DI}
\end{figure*}

\begin{figure*}
\centering
\includegraphics[scale = 0.38]{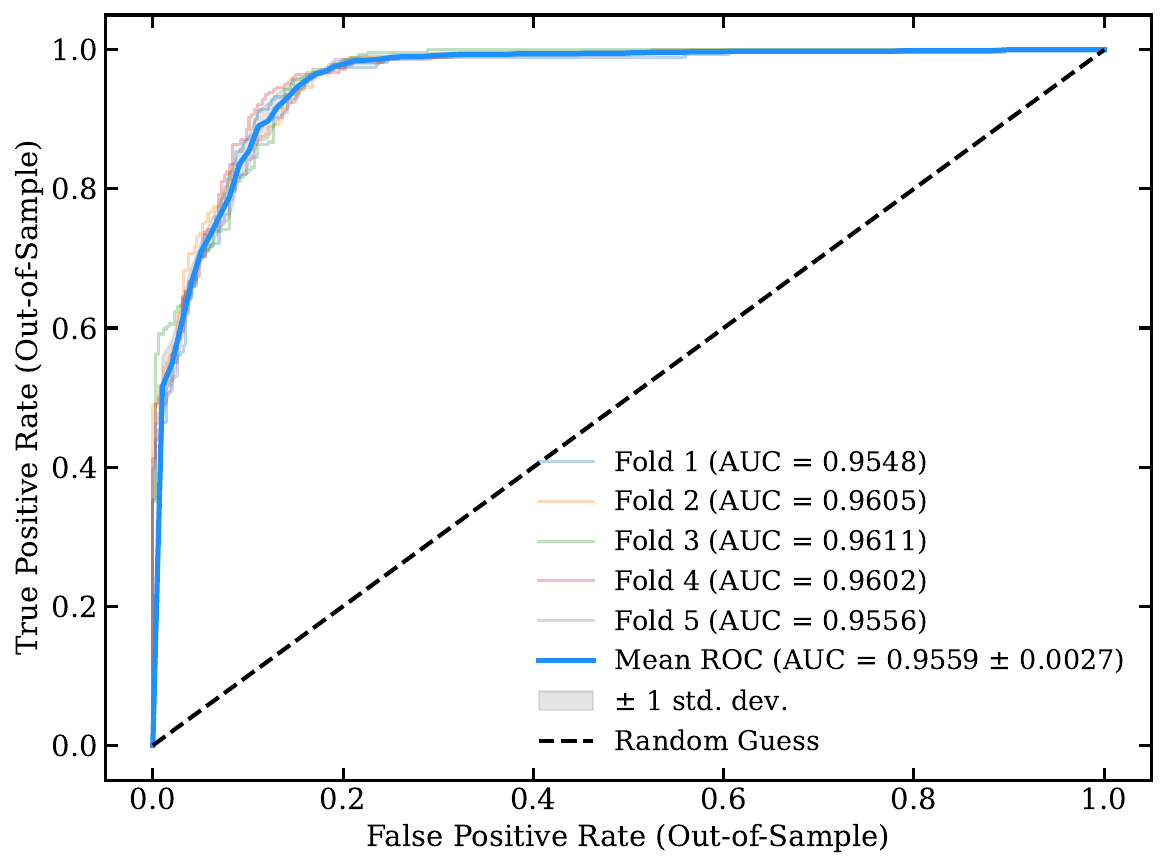}
\includegraphics[scale = 0.38]{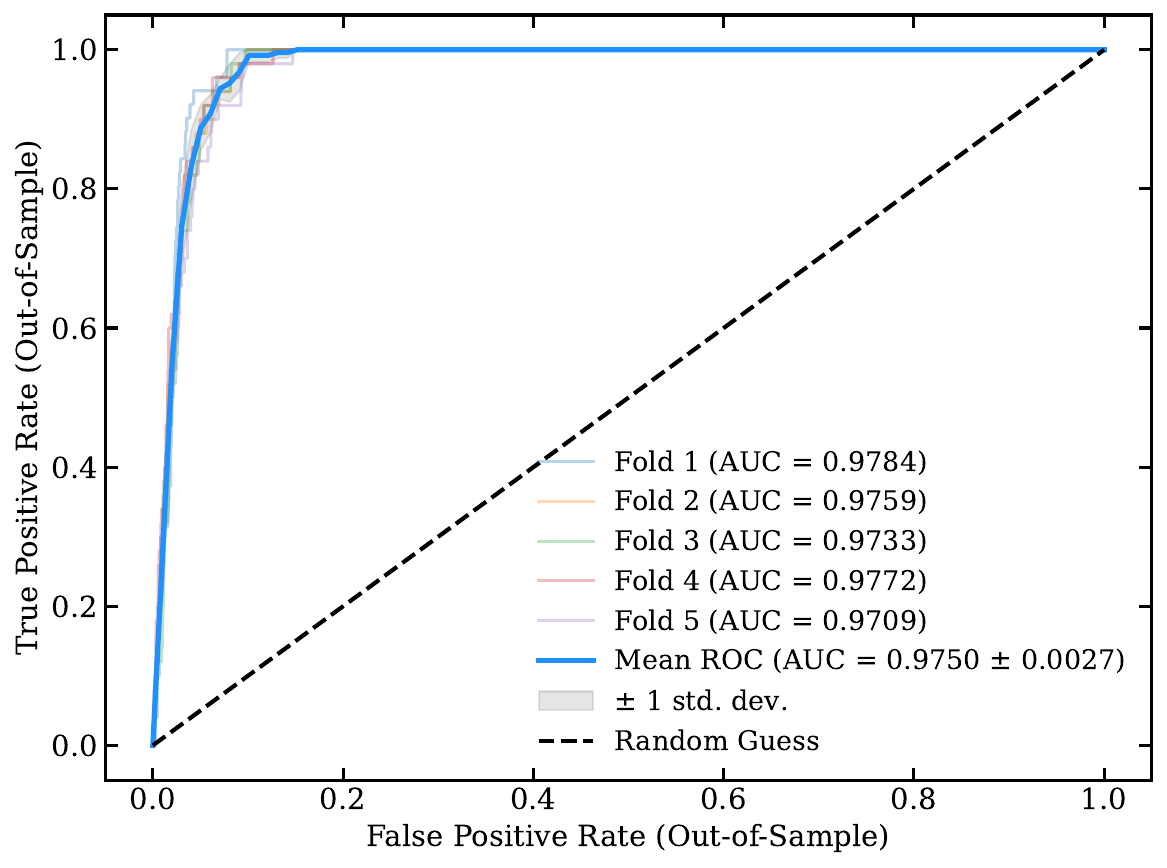}
\caption{ROC curve for the 5-fold stratified cross-validation is plotted for the BNS (\textit{left}) and NSBH (\textit{right}) populations. In each panel, the AUC of the ROC is also shown, along with the ROC for a random classifier.}
\label{Fig:cross_valid_DI}
\end{figure*}

\subsection{Statistics on Detectability Index}
In this section, we discuss the robustness and generalizability of the detectability index ($\rm DI$). To determine the coefficients ($\rm c_{i}$) for $\rm DI_{BNS}$, we train a logistic regression classifier on a population of lensed light curves generated using the BNS-gen model. We frame this as a binary classification problem comprising two classes: (1) GW detectable, and (2) jointly GW and EM detectable (GW+EM). The GW+EM detectable class encompasses all lensed BNS mergers that produce a light curve with a peak magnitude above the single-exposure limit of the $g$-band. The remainder of the population is assigned to the solely GW detectable class. Similarly, to obtain the coefficients ($\rm d_{i}$) for $\rm DI_{NSBH}$, we train a logistic regression classifier on lensed light curves generated by the Bu2019nsbh model, maintaining the identical two-class structure and EM detectability threshold used for $\rm DI_{BNS}$.

A single training iteration on these datasets yields the best-fit coefficients but provides no estimation of variance or associated error margins. To rigorously quantify the stability of the $\rm DI$, we employ a bootstrapping technique. The dataset is resampled with replacement to generate $1,000$ independent bootstrap distributions. Logistic regression optimization is applied to each distribution, yielding a distribution of the fitted coefficients, from which we extract the $90\%$ confidence intervals. We normalize the value of $\rm DI$ such that $\rm c_1 = 1.0$ and $\rm d_1 = 1.0$. The resulting coefficients for both $\rm DI_{BNS}$ and $\rm DI_{NSBH}$, alongside their corresponding $90\%$ confidence intervals, are summarized in Table~\ref{tab:ci_di_coefficients}.

\begin{table}[htbp]
    \centering
    \caption{Best-fit coefficients for the $\rm DI_{BNS}$ ($\rm c_i$) and $\rm DI_{NSBH}$ ($\rm d_i$) models, including $90\%$ confidence intervals derived from $1,000$ bootstrap iterations. The coefficients are normalized such that $\rm c_1, d_1 = 1.0$. The second column lists the quantity whose logarithm each coefficient multiplies.}
    \label{tab:ci_di_coefficients}
    \renewcommand{\arraystretch}{1.3}
    \begin{tabular*}{\columnwidth}{@{\extracolsep{\fill}}cccc@{}}
        \hline
        \hline
        Index & \multirow{2}{*}{Quantity} & $\rm DI_{BNS}$ & $\rm DI_{NSBH}$ \\
        ($i$) &                           & ($\rm c_i$)    & ($\rm d_i$)     \\
        \hline
        1 & $\log(\mu)$          & $1.0^{0.0}_{0.0}$           & $1.0^{0.0}_{0.0}$ \\
        2 & $\log(M)$            & $17.56^{19.79}_{15.65}$     & $-2.70^{-2.26}_{-3.19}$ \\
        3 & $\log(q)$            & $0.24^{0.43}_{0.06}$        & $0.55^{0.78}_{0.43}$ \\
        4 & $\log(|\cos \iota|)$ & $-1.81^{-1.32}_{-2.44}$     & $-4.77^{-4.04}_{-5.06}$ \\
        5 & $\log(D_{L})$        & $-1.60^{-1.53}_{-1.68}$     & $-2.04^{-1.95}_{-2.14}$ \\
        6 & $\log(\chi)$         & ---                         & $4.57^{5.34}_{3.97}$ \\
        \hline
    \end{tabular*}
\end{table}

In Fig.~\ref{Fig:BNS_NSBH_DI}, the detectability index is shown for the lensed BNS and NSBH population. The coefficients $\rm c_i, d_i$ are chosen to be the medians of the bootstrap distribution. We can see that the detectability index separates the GW$+$EM detectable population from the GW only detectable population. From Fig.~\ref{Fig:BNS_NSBH_DI}, it is clear that the higher the value of the detectability index, the higher will be the chance that a GW detectable lensed BNS or NSBH is also observed in EM. 

Further, to test the generalizability of the detectability index, we need to evaluate its performance on unseen data points. To measure the out-of-sample predictive power of the metric, we implement a 5-fold stratified cross-validation. Here, the dataset is partitioned into five mutually exclusive subsets (folds). The stratification ensures that all folds have an identical distribution of both classes. The model is iteratively trained on four folds and evaluated purely on the unseen fifth fold. In Fig.~\ref{Fig:cross_valid_DI}, we show the Receiver Operating Characteristic (ROC) curves and also compute the Area Under the Curve (AUC) for the 5-fold stratified cross-validation of the BNS and NSBH populations. From Fig.~\ref{Fig:cross_valid_DI}, it is clear that the ROC curves on all the folds show a consistent behavior, and the AUC value of the ROCs doesn't show much variance. This high, narrow band of out-of-sample AUC scores confirms that the derived equation is not overfitting the specific training sample but rather accurately predicts the EM detectability of GW-detected lensed BNS and NSBH mergers.

\section{Summary and Conclusion}\label{conclusion_section}
We assess the feasibility of joint detections of lensed GWs and their kilonova counterparts by constructing lensed BNS and NSBH populations for the O5 and LSST observing scenarios.
For the lensed BNS population detectable in the O5 observing scenario, we simulate kilonova light curves across six LSST bands using the semi-analytical BNS-generative model within MOSFiT. Gravitational lensing magnification significantly elevates detectability compared to the largely unobservable unlensed population; for the second and third lensed images, the $g$-band achieves the highest single-exposure detection fraction at $51\%$, which further rises to $58\%$ when incorporating the unresolved image criterion. The joint GW+EM detectable BNS population is dominated by quadruple-image systems with magnifications of $\mathcal{O}(100)$, exhibiting a clear preference for higher total masses and symmetric mass ratios, which generate the bluer, shock-heated ejecta that enhance brightness at shorter optical wavelengths.

For the lensed NSBH population, we model kilonova emission by comparing the NSBH-gen model in MOSFiT against Bu2019nsbh radiative transfer simulations from POSSIS. Formation of counterparts is far more restrictive than in BNS systems: only $\sim 30\%$ of GW-detectable NSBH mergers produce an EM counterpart, requiring high mass ratios and rapidly spinning black holes that prevent a direct plunge of the NS. Furthermore, for systems that do generate light curves, only $ \sim10\%$ are detectable at the LSST single-exposure limit under the generative model, indicating that joint detections of lensed NSBH events will be substantially rarer than lensed BNS events.

To facilitate efficient EM follow-up, we construct a data-driven $\rm DI$ that uses lensed GW parameters to estimate a likelihood of an optical counterpart observable in LSST. We confirm the statistical stability and generalizability of $\rm DI$ via bootstrapping ($90\%$ confidence intervals) and $5$-fold stratified cross-validation, where consistent ROC curves across folds demonstrate strong predictive performance. Although the absolute calibration of $\rm DI$ remains tied to our specific population synthesis assumptions and contemporary kilonova models, it establishes a reliable screening framework that will naturally gain precision as theoretical simulations improve.

This study evaluates the prospects for joint gravitational-wave and kilonova detections of gravitationally lensed binary neutron star and neutron star–black hole mergers during the O5 observing run, while also assessing whether such lensed systems can provide early warnings for upcoming observing scenarios. \cite{2023MNRAS.520..702S} suggested that lensed kilonova originating from BNS mergers could be detected out to a redshift of $z \sim 2$ by the Vera C. Rubin Observatory, primarily due to lensing magnification bias. Consequently, this work serves as a pilot study to evaluate the detectability of EM counterparts within the upcoming LSST. The survey is projected to discover $\mathcal{O}(10^4)$ strongly lensed systems and will likely play a crucial role in following up BNS and NSBH candidates identified by the LVK low-latency alert infrastructure. \cite{2023MNRAS.518.6183M} estimated the rates for the joint detection of lensed BNS mergers in the next-generation GW detectors and EM telescopes. However, their analysis employed a simplistic kilonova model lacking both anisotropy and viewing-angle dependence. In contrast, our study accounts for these critical factors by implementing current benchmark forward models for BNS and NSBH kilonova. While these theoretical models are subject to ongoing refinement, the anticipated increase in future kilonova detections and their associated afterglows will facilitate significant improvements in kilonova modeling.

Furthermore, detectability from the EM side will substantially benefit from the advanced space-based imaging capabilities of the James Webb Space Telescope. Utilizing target-of-opportunity observation modes, it can follow up LVK-triggered BNS and NSBH candidates by surveying galaxies within the GW localization area \citep{Bartos_2016}, thereby enabling the precise modeling of these lens systems. Additionally, next-generation ground-based extremely large telescopes---namely the Thirty Meter Telescope (TMT), the Giant Magellan Telescope (GMT), and the European Extremely Large Telescope (EELT)---will achieve unprecedented deep-field sensitivities ($\sim 27$ mag) and high spatial resolutions ($\sim 0.01$ arcsec). By enabling the detection of faint EM counterparts of lensed BNS and NSBH mergers, these facilities will significantly enhance the potential of lensing-driven early warning systems.


\section*{Acknowledgements}
S.J.K. gratefully acknowledges support from ANRF/SERB grants SRG/2023/000419 and MTR/2023/000086. We acknowledge the use of IUCAA LDG cluster Sarathi for the computational/numerical work. This material is based upon work supported by NSF’s LIGO Laboratory which is a major facility fully funded by the National Science Foundation. 
\vspace{5mm}

\textit{Software}: \texttt{MOSFiT} \citep{2018ApJS..236....6G}, \texttt{POSSIS} \citep{2019MNRAS.489.5037B}, \texttt{NumPy} \citep{vanderWalt:2011bqk}, \texttt{SciPy} \citep{Virtanen:2019joe}, \texttt{astropy} \citep{2013A&A...558A..33A, 2018AJ....156..123A}, \texttt{Matplotlib} \citep{Hunter:2007}, \texttt{jupyter} \citep{jupyter}, \texttt{pycbc}\citep{2018PhRvD..98b4050N}.

\bibliographystyle{aasjournalv7}
\bibliography{references}

\appendix 
We prepare the lensed GW population on BNS and NSBH with the prescription given in \citep{haris2018, Magare_2023}.
We outline a few details here:\\

{1) \textit{Source redshift distribution:} 
We assume that the CBCs are distributed in redshift $z_s$ according to the Madau-Dickinson \citep{Madau:2014bja} distribution:
\begin{equation}
p(z) \propto (1 + z)^{\alpha}\left[1 + \left(\frac{1 +z}{1 + z_p} \right)^{\alpha + \beta}\right]^{-1}
\end{equation}
based on the star formation rate density history. We use the fiducial values of $\alpha = 2.7,~\beta = 2.9, z_p = 1.9$, which correspond to a zero delay time between the formation of the stars in the binary and its merger. \newline The lensing fraction depends on the lensing optical depth $\tau(z_s)$, which is a measure of the probability that the GWs traveling from source to earth will encounter at least one lens. We estimate this probability assuming that the lenses have a constant number density and their velocity dispersion distribution follows early-type galaxies \citep{Sheth_2003}:
\begin{equation}
P(\mathcal{H}_L | z_s) = 1 - \exp(-\tau(z_s)),
\end{equation}
\begin{equation*}
    \tau(z_s) = 4.17\times 10^{-6}\left(\frac{D^c(z_s)}{\mathrm{Gpc}}\right)^3
\end{equation*}

We sample the Madau-Dickinson distribution weighted by the lensing probability, $p(z_s, \mathcal{H}_L) = p(z_s)P(\mathcal{H}_L | z_s)$, to obtain source redshift samples.\\ 

{2) \textit{Distribution of other extrinsic source parameters}:} We assume that the sources are isotropically distributed
 in the sky, which amounts to drawing right ascension and declination samples uniformly in $\alpha \in [0, 2\pi), \cos\delta \in [0, 1]$. 
We also assume that the inclination -- the angle between the orbital angular momentum and the line of sight -- is isotropically distributed on the sphere: $\sin\iota \in [0, 1]$.\\

{3) \textit{Distribution of instrinsic source parameters}:} The BNS component masses ($m_1, m_2$ measured in $M_{\odot}$) are drawn uniformly from $[1.1, 2.1]$, which lies in the typical NS mass range and below the $\rm M_{TOV}$ limit, ensuring that $m_1 > m_2$. The NSBH primary mass is drawn uniformly from $m_1 \in [5, 20]$, and the secondary is drawn uniformly from $m_2 \in [1.1, 2.1]$. The  All binaries are assumed to be non-spinning for the GW scenario and for calculating SNR.\\ 

{4) \textit{Distribution of lens redshifts}: } We define the ratio of comoving distances to the lens and the source as $x \equiv D^c(z_l)/D^c(z_s)$. We then draw this ratio from the polynomial 
\citep{1984ApJ...284....1T}
\begin{equation}
p(x) = 30x^2(1-x)^2
\end{equation} 
The lens redshift samples are then acquired trivially from the ratio samples and the lensed source redshift samples drawn in step 1. The distribution for intrinsic lens parameters and the distribution of the sources in the source plane are according to \citep{Magare_2023}.\\

{5) \textit{GW Observing Scenarios}:} We define the O5 detector network as: 3 LIGO detectors, including LIGO India, at A+ sensitivity \citep{Iyer:2011ligo, 2022CQGra..39b5004S}, 1 Virgo \citep{2015CQGra..32b4001A} and 1 Kagra detector \citep{2021PTEP.2021eA101A}, at target sensitivity. 
See \citep{KAGRA:2013rdx, Hall2019} for additional details on the observing scenarios, and\cite{psds, pycbc} for details on the assumed noise PSDs.\\ 




\end{document}